\documentclass[superscriptaddress,twocolumn,showpacs,preprintnumbers,amsmath,amssymb, prl, reprint, longbibliography, showkeys]{revtex4-1}
\usepackage{amsmath}
\usepackage{amssymb}
\usepackage{graphicx}
\usepackage{stmaryrd}
\usepackage{txfonts}
\usepackage{mathrsfs}
\usepackage{dcolumn}
\usepackage{pbox}
\usepackage{bm}
\usepackage{epsfig}
\usepackage{color}

\usepackage[colorlinks=true, linkcolor=blue, citecolor=blue, urlcolor=blue]{hyperref}
\begin{document}

\title{Impurity-induced magnetic droplet in unconventional superconductors near magnetic instability: Application to Nd doped $\mathbf{CeCoIn_5}$}

\author{Shi-Zeng Lin}
\email{szl@lanl.gov}
\affiliation{Theoretical Division, T-4 and CNLS, Los Alamos National Laboratory, Los Alamos, New Mexico 87545, USA}

\author{Jian-Xin Zhu}
\affiliation{Theoretical Division and Center for Integrated Nanotechnologies, Los Alamos National Laboratory, Los Alamos, New Mexico 87545, USA}

\begin{abstract}
Beside the spin density wave (SDW) order inside the superconducting phase in $\mathrm{CeCoIn_5}$ at high magnetic fields, recent neutron scattering measurements have found Bragg peaks in $5\%$ Nd doped $\mathrm{CeCoIn_5}$ at low fields. The intensity of Bragg peaks in low fields is suppressed by increasing field. Based on the phenomenological and microscopic modeling, we show that for the Pauli limited $d$-wave superconductors in the vicinity of SDW instability relevant for $\mathrm{CeCoIn_5}$, magnetic impurities locally induce droplets of SDW order. Because of the strong anisotropy in the momentum space in the spin fluctuations guaranteed by the $d$-wave pairing symmetry, sharp peaks in spin structure factor at $\mathbf{Q}$s are produced by the impurities, even when the droplets of SDW do not order. At zero field, the Nd impurity spins are along the $c$ axis due to the coupling to the conduction electrons with an easy $c$ axis, besides their own crystal field effect. The in-plane magnetic field cants the impurity moments toward the $ab$ plane, which suppress the droplets of SDW order. At high fields, the long-range SDW inside the superconducting phase is stabilized as a consequence of magnon condensation. Our results are consistent with the recent neutron scattering and thermal conductivity measurements.

\end{abstract}

\date{\today}
\maketitle

\noindent
\textbf{Introduction} --
The intricate interplay between magnetism and superconductivity represents one of main challenges in strongly correlated electronic systems. The $d$-wave heavy-fermion superconductor $\mathrm{CeCoIn_5}$ is one prototypical system \cite{petrovic_heavy-fermion_2001,thompson_progress_2011} to study the relationship between magnetism and superconductivity in Pauli-limited superconductors. Conventionally, superconductivity is stabilized through suppression of magnetism by pressure, chemical doping etc. The discovery that in $\mathrm{CeCoIn_5}$ a spin density wave (SDW) emerges only inside the superconducting phase by external field of the order of $10\  \mathrm{T}$ comes as a big surprise \cite{kenzelmann_coupled_2008,PhysRevLett.104.127001,PhysRevLett.98.036402}. This experiment immediately triggers enormous experimental and theoretical efforts to understand the origin of the SDW phase and the role of superconductivity. Neutron scattering studies revealed that the propagation wave vector is $\mathbf{Q}=(\pm 0.44,\ \pm 0.44,\ 0.5)$, which coincides with the nodal direction of $d_{x^2- y^2}$ pairing symmetry. The fully developed magnetic moment is $0.15\ \mu_B$ and the magnetic moment is along the $c$ axis. Here $\mu_B$ is the Bohr magneton. The ordering wave vector of SDW along the two perpendicular nodal direction can be switched sensitively by rotating the in-plane magnetic field \cite{gerber_switching_2014,Kim2016}. The neutron study in the superconducting phase at low fields has revealed plenty of magnetic fluctuations of nodal quasiparticle at the same $\mathbf{Q}$ and spin anisotropy as those in the SDW phase \cite{PhysRevLett.100.087001,PhysRevLett.115.037001,PhysRevLett.109.167207}. This implies that the SDW is due to the condensation of these magnetic fluctuatios.

Meanwhile several theoretical proposals have been put forward to understand the origin of high field SDW phase. These theories highlight the importance of vortex lattice \cite{PhysRevB.83.140503}, Pauli pair breaking effect \cite{PhysRevB.82.060510,PhysRevB.83.224518}, the Fulde-Ferrell-Larkin-Ovchinnikov (FFLO) state, superconducting pairing density wave \cite{PhysRevLett.102.207004,yanase_antiferromagnetic_2009,miyake_theory_2008,PhysRevLett.104.216403}, and improved Fermi surface nesting by magnetic field \cite{PhysRevLett.107.096401,PhysRevB.86.174517,PhysRevB.92.224510}, in stabilizing the SDW phase. Using the tight-binding parameters obtained by density functional calculations with renormalized bandwidth, it was shown that the in-plane magnetic field enhances the transverse magnetic susceptibility, $\chi(\mathbf{q}, \omega)$ in the Pauli-limited $d$-wave superconductors \cite{PhysRevB.84.052508}. Using the argument based on the random phase approximation (RPA), the enhanced susceptibility triggers the SDW instability when $U(\mathbf{Q})\chi(\mathbf{Q})= 1$, where $U(\mathbf{q})$ is the Coulomb interaction. In these theories, it is assumed that $\mathrm{CeCoIn_5}$ is in the vicinity of the SDW instability, which can be inferred from various measurements \cite{PhysRevLett.91.257001,PhysRevLett.91.246405,PhysRevB.73.064519,PhysRevLett.108.056401,doi:10.1143/JPSJ.72.2308,PhysRevLett.100.087001,PhysRevLett.115.037001,PhysRevLett.109.167207}.

It is well known that the response of superconductors to impurity is a powerful signature to understand the pairing mechanism in unconventional superconductors \cite{RevModPhys.78.373,PhysRevB.64.140501,PhysRevLett.89.217002,PhysRevLett.89.067003,PhysRevLett.92.077203,PhysRevB.75.054520,PhysRevLett.99.147002,PhysRevLett.105.147002,PhysRevLett.113.067002,PhysRevLett.117.257002}. Previously work on Cd doped $\mathrm{CeCoIn_5}$, $\mathrm{CeCo(In_{1-x}Cd_x)_5}$, has revealed interesting interplay between magnetism and superconductivity \cite{PhysRevB.76.052401,PhysRevLett.97.056404,PhysRevLett.99.146402,nair_magnetism_2010,Seo_NatPhys_2014}. Recently, an anomaly in specific heat has been observed in the superconductor $\mathrm{Nd_xCe_{1-x}CoIn_5}$ for $x=0.1,\ 0.05$ \cite{PhysRevB.77.165129}. The compound with $x=0.05$ has been studied by neutron scattering recently \cite{raymond_magnetic_2013}, where Bragg peaks at the same SDW ordering wave vector, $\mathbf{Q}$, have been observed at zero magnetic field. The magnon fluctuations are gapped with a gap of $0.43\ \mathrm{meV}$. \cite{mazzone_spin_2017} Further neutron studies find that the Bragg peaks at four $\mathbf{Q}$s have the same intensity \cite{KenzelmannPreprint, KenzelmannPreprint2}. The peak intensity in the low-field  region is suppressed by increasing the in-plane magnetic field and eventually disappears before the high-field SDW phase emerges. Thermal conductivity measurement on the $5\%$ Nd doped $\mathrm{CeCoIn_5}$ shows a jump in conductivity  at high magnetic fields when the field is rotated along the $a$ axis, similar to that in the undoped $\mathrm{CeCoIn_5}$ \cite{Kim2016}, suggesting a long-range SDW phase. The jump in thermal conductivity disappears at low magnetic field, indicating a different magnetic response of distinct origin \cite{KimPreprint}. 

Assuming that the $5\%$ Nd doping has negligible effect on the Fermi surface, and the Nd atoms serve as magnetic impurities in the $d$-wave superconductivity, a minimal model based on a quasi-nested Fermi surface was proposed to understand these experiments, where the magnetic impurity induces a locally SDW pattern at low field \cite{PhysRevB.92.224510}. It is argued that these local SDW patterns cooperate to form a long-range SDW order, as a consequence of the Ruderman-Kittel-Kasuya-Yosida (RKKY)-like interaction between magnetic impurities. The phase region of the high-field SDW phase becomes wider in the presence of magnetic impurities. However it is unclear how the low-field SDW region is suppressed by magnetic field.

In this work, we present a theory on the magnetic properties in Nd doped $\mathrm{CeCoIn_5}$. Our theory are based on the following experimentally established facts: (1) the system is close to the SDW instability; (2) the SDW at high fields is formed by the nodal quasiparticles and the ordering wavevector is along the nodal directions of the $d$-wave superconductivity; (3) the moment of SDW order is along the $c$ axis. Within the phenomenological and microscopic modeling, we show that magnetic impurities generate local droplets of SDW oscillation at low fields, which oscillate and decay in space. Because of the $d$-wave pairing symmetry, the random impurities can produce peaks in the spin structure factor, even when they do not order magnetically. The impurity moments experience an easy axis anisotropy due to the conduction electrons, besides the anisotropy generated by crystal field. When an in-plane magnetic field is applied, the impurity moments are canted toward the $ab$-plane, accompanying the reduction of induced local SDW order. At high field, the long-range SDW sets in.

\noindent
\textbf{Phenomenological model} --
Here we present the physical picture based on a phenomenological model. We consider a paramagnetic region inside the superconducting phase, where we keep the free energy expansion of magnetization density $M_z$ up to the quadratic order. It is sufficient to consider the magnetization inside one layer because of antiferromagnetic order between layers. The total free energy density is \cite{KimPreprint2}
\begin{align}\label{eq1}
\begin{split}
{\cal F} = - \frac{\alpha (T, H_x) }{2}M_z^2  - \gamma {\left( {{\nabla _{2d}}{M_z}} \right)^2} \\
+ \eta \left[ {{{\left( {\partial _x^2{M_z}} \right)}^2} + {{\left( {\partial _y^2{M_z}} \right)}^2}} \right]
-J_s S_z M_z-  H_x S_x+\mathcal{H}_{\mathrm{Ani}}(\mathbf{S}),
\end{split}
\end{align}
where $\nabla_{2d}\equiv(\partial_x,\ \partial_y)$, $\mathbf{S}$ is the impurity moment and $H_x$ is the in-plane magnetic field. Here $\mathcal{H}_{\mathrm{Ani}}(\mathbf{S})$ is the spin anisotropy for the impurities, which arises from their coupling to the conduction electrons with easy axis anisotropy and crystal field environment. For $\gamma>0$, $\eta>0$, the $\mathbf{Q}$ of the SDW is $Q_x=\pm Q_y=\pm\sqrt{\gamma/2\eta}$. We have expanded the anisotropy of $\mathbf{Q}(\theta)$ in the $ab$ plane up to $\cos (4\theta)$ represented by the $\eta$ term, where $\theta$ is the angle between $\mathbf{Q}$ and the crystal $a$ axis. For a stronger anisotropy, one needs to include higher order expansion in $\theta$. We have neglected the spin-orbit interaction of the conduction electrons that is responsible for the switching of $\mathbf{Q}$ of the SDW order by rotating magnetic field \cite{KimPreprint2}. 

That the system is close to the SDW instability means that the magnetic susceptibility 
\begin{align}\label{eq2}
\chi(\mathbf{q})=\frac{1}{ - \alpha  - 2\gamma {q^2} + 2\eta \left( {q_x^4 + q_y^4} \right)},
\end{align} 
is positive and close to divergence. Microscopic theories show that $\alpha(T, H_x)$ increases with $H_x$. At a critical field, $\chi(\mathbf{Q})$ diverges and the magnons condense to form the SDW.

\begin{figure}[t]
\psfig{figure=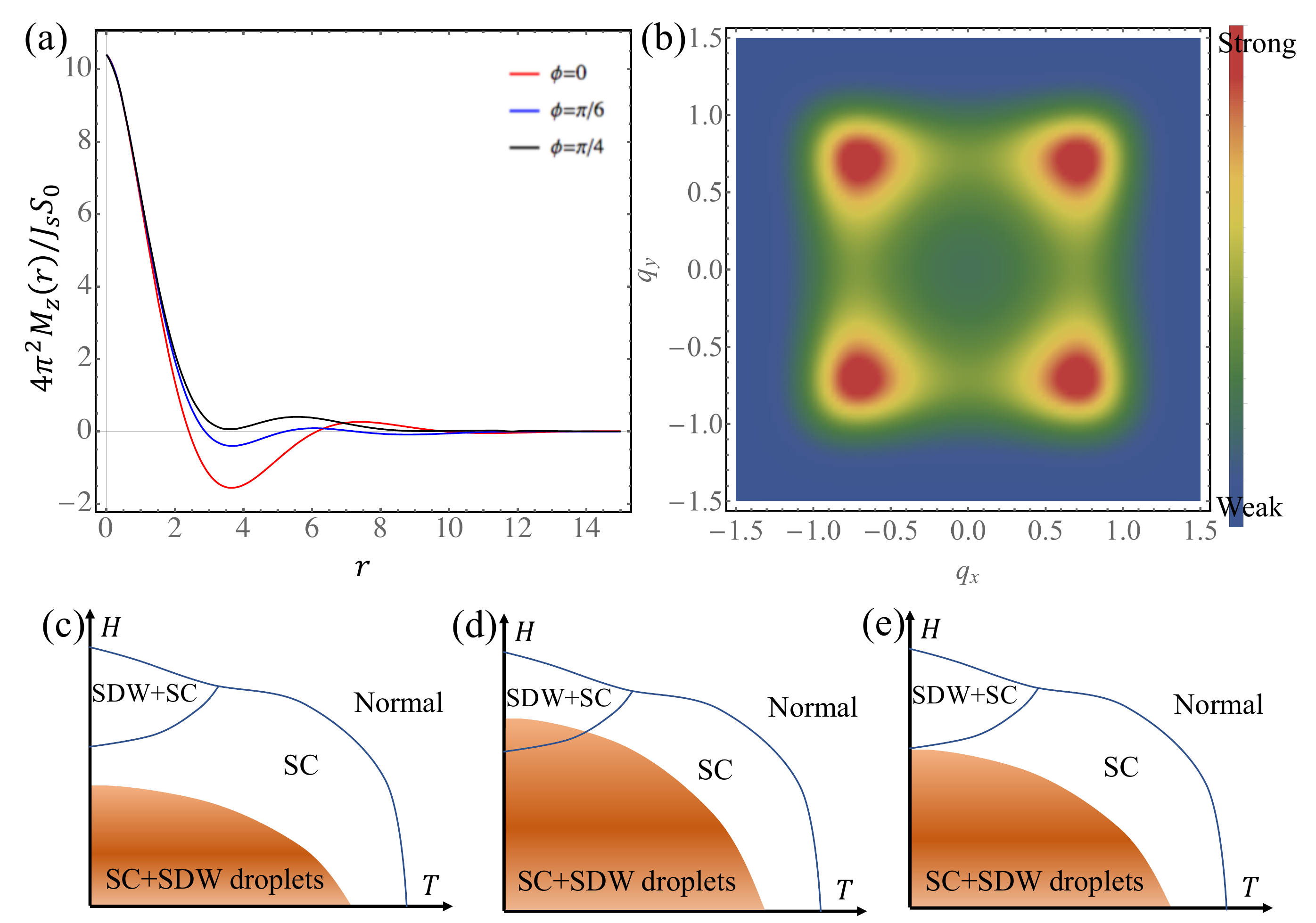,width=\columnwidth}
\caption{(color online) (a) Magnetization profile for different angle $\phi$ between $\mathbf{r}$ and the $x$ axis according to Eq. \eqref{eq3}.  (b) Impurity-induced spin structure factor. Here $\gamma=\eta=1/2$ and $\alpha=1$. (c-e) Schematic view of the phase diagram for $d$-wave superconductor (SC) at the brink of SDW instability doped by dilute magnetic impurities. In the shaded region, droplets of local SDW are induced by impurities.
} \label{f1}
\end{figure}

For a single impurity located at the origin, $S_z=S_0 \delta(\mathbf{r})$, the induced local SDW pattern at zero field is
\begin{align}\label{eq3}
{M_z}\left( \mathbf{r} \right) = \frac{J_s S_0}{{{{\left( {2\pi } \right)}^2}}}\int\frac{{\exp \left( {{{i}}{\bf{q}}\cdot{\bf{r}}} \right)}}{{ - \alpha  - 2\gamma {q^2} + 2\eta \left( {q_x^4 + q_y^4} \right)}}d{\mathbf{q}^2}.
\end{align}
Here $M_z(\mathbf{r})$ oscillates and then decays as a function of distance as depicted in Fig. \ref{f1} (a), where both the oscillation period and decay length depend on the angle between $\mathbf{r}$ and the $x$ axis. The decay is slow because the $\chi(\mathbf{Q})$ is close to divergence.

We then consider random distribution of impurities with the correlation $\langle S_z \rangle=0$ and $\langle S_z(\mathbf{r})S_z(\mathbf{r}') \rangle=A(H_x) \delta(\mathbf{r}-\mathbf{r}')$.
We introduce an effective saturation field $H_s$ to describe the canting of impurity spin by magnetic field. Therefore, we have $A=A_0(1-H_x^2/H_s^2)$ when $H_x\le H_s$ and $A=0$ otherwise. A self-consistent treatment of the impurity moment direction will be presented in the microscopic model. The spin structure factor $\mathcal{S}_{zz}(\mathbf{q})=\langle M_z(\mathbf{q})M_z(-\mathbf{q})\rangle$ is
\begin{align}\label{eq4}
\mathcal{S}_{zz}(\mathbf{q})=\frac{A J_s^2}{[{ - \alpha(T, H_x)  - 2\gamma {q^2} + 2\eta ( {q_x^4 + q_y^4})}]^2}.
\end{align} 
The impurity induced $\mathcal{S}_{zz}(\mathbf{q})$ is maximal at four $\mathbf{Q}$s, $(\pm Q_x,\ \pm Q_y)$, with equal intensity, because of the strong anisotropy in the orientation of $\mathbf{Q}$, see Fig. \ref{f1} (b), even for randomly distributed impurities. In $\mathrm{CeCoIn_5}$, $\chi(\mathbf{q})$ increases with magnetic field while $A$ decreases with field. Depending on the relative strength of these two competing factors, $\mathcal{S}_{zz}(\mathbf{Q})$ can increase or decrease monotonically, or show nonmonotonic behavior as a function of $H_x$ in the paramagnetic phase. Taking the high field SDW phase into account, we can schematically draw the magnetic response of the Pauli limited $d$-wave superconducting phase in the vicinity of SDW to magnetic impurities, as shown in Fig. \ref{f1} (c-e). Because the system is close to the SDW instability, the induced SDW droplets decay slowly in space, which may be helpful to achieve a long-range order of the SDW droplets \cite{PhysRevLett.117.257002,PhysRevB.92.224510,PhysRevLett.113.067002}. The low-field phase can overlap with the long-range SDW phase or they can separate, or accidentally touch the SDW phase at a point, depending on the saturation field for the impurities.

\noindent
\textbf{Microscopic model} --
We present microscopic model calculations to support the phenomenological model. The mean-field Hamiltonian reads
\begin{align}\label{eq6}
\begin{split}
{\cal H} = \sum_{i, j,\sigma}t_{ij}c_{i\sigma}^{\dagger}c_{j\sigma}-\mu\sum_{i,\sigma}c_{i\sigma}^{\dagger}c_{i\sigma}+\sum_{\langle i, j\rangle }(\Delta_{ij}c_{i\uparrow}^{\dagger}c_{j\downarrow}^{\dagger}+\Delta_{ij}^*c_{j\downarrow}c_{i\uparrow})\\
-\sum_{i}\left(h_{i} s_{z, i}-\mu_c H_x s_{x, i}\right)- \sum_{j\in\mathrm{impurity}}\left( J_S S_{z, j} s_{z,j}-\mu_S H_x S_{x, j}\right).
\end{split}
\end{align}
where the spin of the conduction electrons is $\mathbf{s}_{i}=\sum_{\alpha'\beta'}c_{i\alpha'}^{\dagger}\mathbf{\sigma}_{\alpha'\beta'}c_{i\beta'}$. $\mu_c$ and $\mu_S$ are the magnetic moments for the conduction electrons and impurity spins. The self-consistent condition for the anisotropic molecular field $h_i$ and pairing potential $\Delta_{ij}$ are $h_i=-\sum_{j} J_{ij}(n_{j\uparrow}-n_{j\downarrow})$ and $\Delta_{ij}=\frac{V}{4}\sum_l\left(u_{i\uparrow,l}v_{j\downarrow,l}^*+u_{j\uparrow,l}v_{i\downarrow,l}^*\right)\tanh(E_l/2T)$. The electron density for up and down spin are $n_{i\uparrow}=\sum_l|u_{i\uparrow,l}| f(E_l)$ and $n_{i\downarrow}=\sum_l |v_{i\downarrow,l}| f(-E_l)$. Here $u_{i\uparrow,l}$, $v_{i\downarrow,l}$, and $E_l$ are the $l$-th eigen vector and eigen energy of the Bogoliubov-de Gennes equation associated with Eq. \eqref{eq6} \cite{JXZhuBook}. We consider electron hopping on a square lattice with the dispersion $\epsilon(\mathbf{k})=2 t_1 [\cos(k_x)+\cos(k_y)]+4 t_2\cos(k_x)\cos(k_y)+2t_3[\cos(2k_x)+\cos(2k_y)]-\mu$, where $t_1$, $t_2$ and $t_3$ are the nearest neighbor (NN), the second NN (along diagonal) and the third NN (along bond) hopping amplitude respectively. To ensure the SDW order at $\mathbf{Q}$, we use $J_{ij}=J_1$ (NN antiferromagnetic interaction) and $J_3$ (third NN interaction) competing interactions with $J_3=-J_1/4\cos(2\pi Q_x)$. We take the band structure obtained by DFT calculations \cite{tanaka_theory_2006} $t_2=-0.5 t_1$, $t_3=-0.4 t_1$ and the average electronic occupation is fixed at $\langle n\rangle=0.72$. To stabilize the $d$-wave pairing symmetry, we focus on the NN pairing potential $\Delta_{ij}$ with $V=4.5 t_1$ in the calculations. The $d$-wave order parameter is given by $\Delta_d=(\Delta_{i,i+\hat{x}}+\Delta_{i, i-\hat{x}}-\Delta_{i,i+\hat{y}}-\Delta_{i, i-\hat{y}})/4$, where $\hat{x}$ and $\hat{y}$ are the unit vectors in the $x$ and $y$ direction respectively \cite{PhysRevB.50.13883}. We neglect the crystal field effect for the impurities, and the spin anisotropy for the magnetic impurities is only contributed from the conduction electrons. We calculate self-consistently $S_{z, j}$ by fixing $|\mathbf{S}_j|=1$. We use the periodic boundary condition and the system size is chosen to be commensurate with the wavelength of SDW. 

\begin{figure}[t]
\psfig{figure=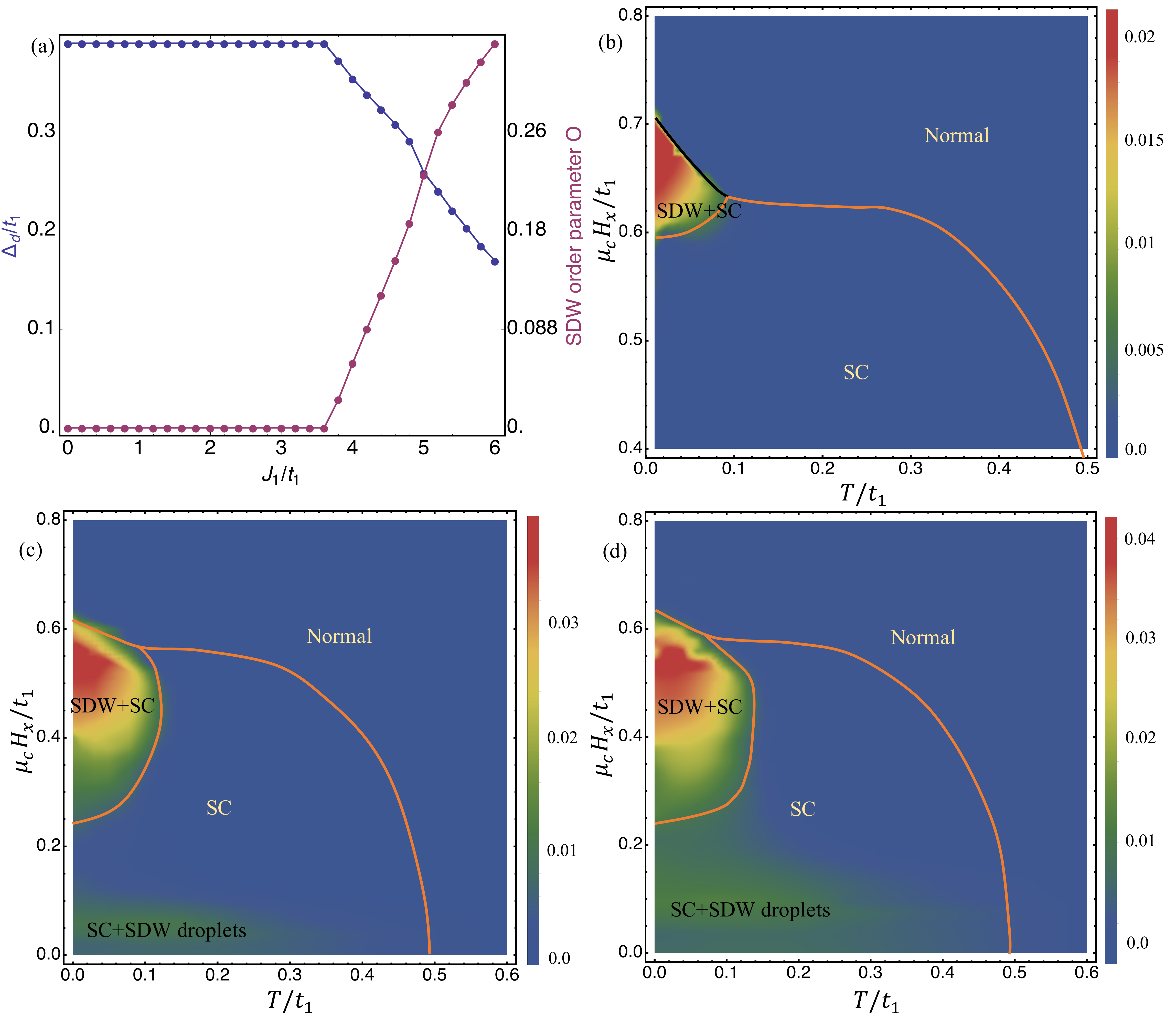,width=\columnwidth}
\caption{(color online)  (a) SDW and $d$-wave order parameter as a function of $J_1$ at zero temperature. Temperature-magnetic field phase diagram for  (b) clean system, (c) system with impurities with $\mu_S/t_1=8$ (lower saturation field), and (d) system with impurities with $\mu_S/t_1=5$ (higher saturation field). Orange (black) lines denote the second (first) order phase transition. The impurity concentration is $5\%$. The magnetization induced by impurities in the normal state has been subtracted.
} \label{f2}
\end{figure}

As shown in Fig. \ref{f2}(a), for a large $J_1$, SDW order emerges and coexists with superconductivity. We tune the system close to the SDW instability by choosing $J_1=3.2 t_1$, close to the critical value for the onset of the SDW order. For a clean system, the transverse susceptibility increases with $H_x$. At high field, a SDW phase inside the superconducting phase with $\mathbf{Q}$ either along $\mathbf{Q}_1=(0.44,\ 0.44)$ or $\mathbf{Q}_2=(0.44,\ -0.44)$ is stabilized, see Fig. \ref{f2}(b). The transition between the SDW phase and normal phase is of the first order, while the other phase transitions are of the second order. Our self-consistent calculations thus confirm the expectation from the RPA argument in Ref. \onlinecite{PhysRevB.84.052508}.  To obtain the experimental phase diagram, it is required that $V$, $\mu_c H_x$ and $t_1$ are in the same order of magnitude, which highlights the uniqueness of heavy-fermion superconductors for the observation of the field-included SDW order.

\begin{figure}[t]
\psfig{figure=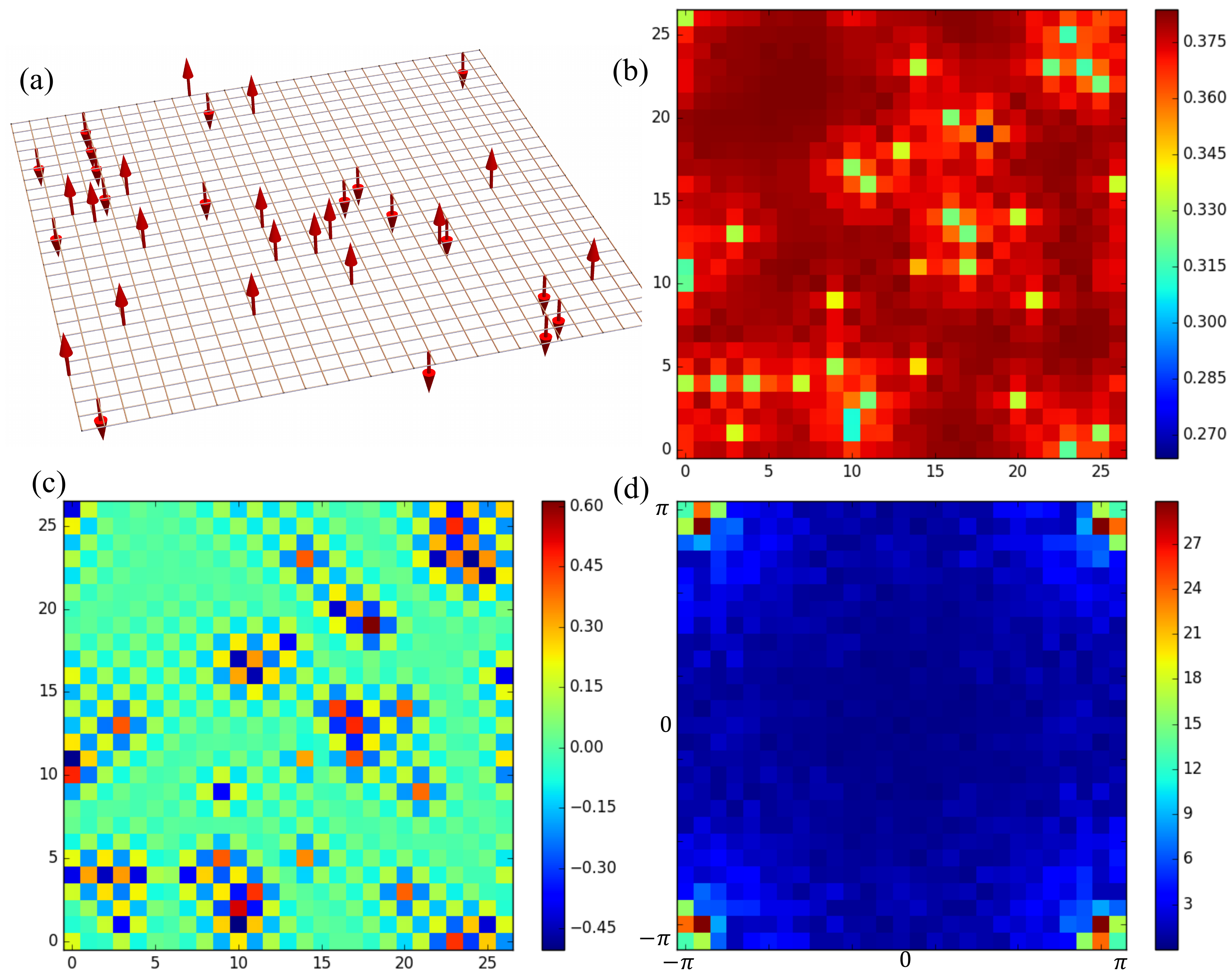,width=\columnwidth}
\caption{(color online) Configurations of (a) impurity spins, (b) amplitude of the $d$-wave order parameter, (c) droplets of SDW, and (d) spin structure factor obtained after averaging over random distribution of impurities. Here $\mu_S=8t_1$, $H_x=0$ and $T=0.01t_1$.} \label{f3}
\end{figure}

We then introduce $5\%$ randomly distributed magnetic impurities into the system. For the impurity coupling $J_S=1.0 t_1$, as shown in 
Fig. \ref{f3}(a), the impurity moment is along the $c$ axis, due to the anisotropic RKKY interaction mediated by conduction electrons with moments in the $c$ axis. The dilute impurity moments do not develop long-range order because of the weak RKKY interaction. The $d$-wave order parameter [see Fig. \ref{f3}(b)] and the superconducting transition temperature $T_c$ is weakly suppressed. Nevertheless, the impurities nucleate droplets of SDW order, which interfere with each other, see Fig. \ref{f3}(c). The spin structure clearly develops Bragg peaks, which preserve the $C4$ rotation symmetry because of the strong anisotropy in $\mathbf{Q}$s, as displayed in Fig. \ref{f3}(d), even though the magnetic impurities do not order.

The fully self-consistent calculations enable us to construct field-temperature phase diagrams in the presence of magnetic impurities.  In Fig. \ref{f2}(c-d), we depict the phase diagram obtained by $\Delta_d$ and SDW order parameter $O \equiv \sum_is_{z,i}^2/N_L$ with $N_L$ the number of lattice sites. There is a long-range SDW phase at high field and low temperature. Compared to the clean system, the upper critical field $H_p$ is suppressed by impurities, while the the phase region for the SDW phase increases, because of the suppression of superconductivity by impurities, which in turn favors its competitor, the SDW phase. The strong first-order phase transition between the SDW phase and normal phase is weakened to the second order in our two-dimensional model. 
In three dimensions, the transition could remain to be first order. At low fields, the impurities induce droplets of local SDW. The induced magnetization decreases with the magnetic field because of the canting of impurity moments in the presence of the field. For a weak saturation field, the region with induced SDW droplet in the phase diagram is separated from the long-range SDW phase [Fig. \ref{f2}(c)]. When the saturation field is increased, these two regions overlap [Fig. \ref{f2}(d)]. The phase diagram is consistent with that expected from the phenomenological model.

\noindent
\textbf{Discussions} -- The amplitude of induced droplet of SDW is proportional to the impurity moment along the easy axis of the conduction electron spins. For magnetic impurity with strong easy-plane anisotropy in the tetragonal crystal, such as Gd, the local exchange interaction between conduction electron spins becomes negligible, therefore no droplets of SDW would be produced. This is consistent with the recent experiment on Gd doped $\mathrm{CeCoIn_5}$, where the impurity induced anomaly in specific heat observed in Nd doped $\mathrm{CeCoIn}_5$ becomes extremely weak for the Gd doped compound \cite{RosaPreprint}. To induce SDW droplets, one may cant the Gd moments toward the easy axis by applying magnetic field in the $c$ axis. The long-range high-field SDW order does not emerge for this field orientation \cite{PhysRevLett.105.187001}, and the magnetization is due to the SDW droplets. 

We have neglected the spin-orbit interaction. As a consequence, the spin structure factor associated with droplets of SDW have four sharp peaks with equal intensity. The spin-orbit interaction couples $\mathbf{Q}$ of magnon fluctuations to the in-plane magnetic field \cite{MineevSwitching,KimPreprint2}, therefore favors a pair of $\mathbf{Q}$s along the diagonal that is as perpendicular to the magnetic field as possible over the other pair for a nonzero field.

To summarize, based on both phenomenological and microscopic models, we have provided a theoretical understanding of the effects of magnetic impurity in unconventional Pauli limited superconductors in the brink of SDW instability, and have applied it to the Nd doped $\mathrm{CeCoIn_5}$. Impurities with magnetic moment parallel to that of conduction electron spins locally induce droplets of SDW, whose amplitude decays in space. Even when these droplets do not order, the spin structure factor exhibits sharp peaks at the same $\mathbf{Q}$s as those of the long-range SDW at high field, because of the strong anisotropy in $\mathbf{Q}$ enforced by the $d$-wave pairing symmetry. With increasing in-plane magnetic field, the impurity moments are canted and the amplitude of the droplets of SDW decreases. At high fields, the long-range SDW inside the superconducting phase is stabilized as a consequence of magnon condensation. Our results are consistent with the recent neutron scattering data and thermal conductivity measurements. The SDW droplets can be probed by NMR measurements.

\begin{acknowledgments}
SZL is indebted to Roman Movshovich, Duk Y. Kim, Priscila F. S. Rosa and Yuan-Yen Tai for helpful discussions. The work was carried out under the auspices of the U.S. DOE Contract No. DE-AC52-06NA25396 through the LDRD program, and supported in part by the Center for Integrated Nanotechnologies, a U.S. DOE Office of Basic Energy Sciences user facility.
\end{acknowledgments}

\bibliography{reference}

\end{document}